\documentclass[11pt]{article}

\usepackage[margin=1in]{geometry}

\usepackage{booktabs}       
\usepackage{graphicx}       
\usepackage{amsmath}        
\usepackage{url}            
\usepackage{microtype}      
\usepackage{multirow}       
\usepackage{caption}        
\usepackage[numbers]{natbib}
\usepackage{hyperref}       
\usepackage{xcolor}         

\hypersetup{
  colorlinks=true,
  linkcolor=blue!70!black,
  citecolor=blue!70!black,
  urlcolor=blue!70!black,
  bookmarksnumbered=true,
  pdfauthor={Rong Lu, Hao Liu, Song Hou},
  pdftitle={Evaluation of Embedding-Based and Generative Methods for LLM-Driven Document Classification: Opportunities and Challenges},
}

\title{Evaluation of Embedding-Based and Generative Methods for\\LLM-Driven Document Classification:\\Opportunities and Challenges}

\author{
  Rong Lu\thanks{Corresponding author.} \quad
  Hao Liu \quad
  Song Hou \\[6pt]
}

\date{}  

\begin{document}

\maketitle

\begin{center}
\small
\textit{Accepted at the IMAGE'25 Workshop (PCW-11), Society of Exploration Geophysicists (SEG).}\\
\textit{Published version available at:} \url{https://doi.org/10.1190/image2025-w11-03.1}
\end{center}

\vspace{1em}

\begin{abstract}
\noindent
This work presents a comparative analysis of embedding-based and generative models for classifying geoscience technical documents. Using a multi-disciplinary benchmark dataset, we evaluated the trade-offs between model accuracy, stability, and computational cost. We find that generative Vision-Language Models (VLMs) like Qwen2.5-VL, enhanced with Chain-of-Thought (CoT) prompting, achieve superior zero-shot accuracy (82\%) compared to state-of-the-art multimodal embedding models like QQMM (63\%). We also demonstrate that while supervised fine-tuning (SFT) can improve VLM performance, it is sensitive to training data imbalance.
\end{abstract}

\section{Introduction}
The oil and gas industry is experiencing an unprecedented data deluge. A vast and growing corpus of technical information resides in unstructured archives of reports, logs, and surveys. Manual classification of these assets is a significant operational bottleneck, making robust, automated systems essential for transforming these archives into actionable intelligence.

A key challenge is the multimodal nature of these documents. Critical classification cues often lie in visual elements such as log charts, seismic sections, and specific page layouts, which are missed by text-based models. Furthermore, the quality of Optical Character Recognition (OCR) can be low on legacy scanned documents, diminishing the effectiveness of text-only approaches. This necessitates the use of models that can jointly process visual and textual information.

Two primary paradigms have emerged: \textit{embedding-based} methods, which generate dense vector representations for similarity-based classification, and \textit{generative} methods, where VLMs directly produce a class label. In this paper, we conducted a comparative study of these two approaches on a benchmark dataset of multi-disciplinary geoscience documents. We investigated the impact of prompting, fine-tuning, and data characteristics, offering insights for practitioners aiming to deploy these technologies effectively.

\section{Methodology}
Our method encompasses a proprietary dataset, a suite of evaluation metrics, and standardized workflows for each modeling paradigm.

\subsection{Dataset}
We curated a benchmark dataset from an internal collection of technical documents. The dataset comprises eight classes spanning key disciplines in the energy sector: Geology \& Geochemistry, Petrophysics, Geophysics, and Petroleum Engineering. The documents are in various formats, including multi-page PDFs and raster images (TIFF/TIF, PNG, JPG). For all experiments, the first page of each document was used to ensure a consistent evaluation basis.

\subsection{Metrics}
We measure classification performance using overall accuracy and macro F1-score. For embedding models, we also evaluate the clustering quality of the ground-truth classes using metrics inspired by \citet{DeBrabandere2017}, i.e., an intra-cluster distance ($L_{\text{intra}}$) to measure cohesion and an inter-cluster distance ($L_{\text{inter}}$) to measure separation:
\begin{align}
L_{\text{intra}} &= \frac{1}{C}\sum_{c=1}^{C} \frac{1}{N_c} \sum_{i=1}^{N_c} d_{\text{cosine}}(\mu_c, x_i) \\
L_{\text{inter}} &= \frac{1}{C(C-1)}\sum_{c_A \neq c_B} d_{\text{cosine}}(\mu_{c_A}, \mu_{c_B})
\end{align}
where $\mu_c$ represents class centroids and $d_{\text{cosine}}$ denotes cosine distance (defined as 1.0 minus the cosine similarity). We also compute the ratio of separation over cohesion, silhouette score, Davies-Bouldin (DB) Index, and Calinski-Harabasz (CH) Index.

\subsection{Similarity-Voting using Embedding}
This approach reframes classification as a similarity-based voting. First, documents are converted to PIL images. Extremely large images are resized to a maximum dimension of 8192 pixels while preserving aspect ratio. Next, both the document images and the class labels are converted to embeddings. For document images, we use a simple prompt that instructs the model to generate a vector representation of the document. For class labels, we found that providing detailed domain-specific definitions boosts performance over using just the class name. Finally, the cosine similarity between the document embedding and each of the class embeddings is calculated. The class with the highest similarity score is chosen as the predicted label. We benchmarked five publicly available multimodal models.

\subsection{VLM with Prompt Engineering}
This approach uses a VLM to directly generate the class label. We designed an advanced prompting strategy that combines CoT reasoning~\citep{10.5555/3600270.3602070} with domain knowledge. The prompt (termed ``plus'' version) instructs the VLM to follow a multi-step process for more robust and accurate reasoning. It significantly improved performance over simpler prompts (termed ``base'' version which adds personas in prompts). The prompt and image are sent to a locally deployed endpoint and the model's generated output is parsed to extract the predicted class. We evaluated four state-of-the-art open-weight VLMs.

\subsection{VLM with SFT}
To evaluate the impact of domain adaptation, we fine-tuned a Qwen2.5-VL-7B model~\citep{2025arXiv250213923B} using around 7000 training samples. The dataset is imbalanced: most classes contain hundreds to thousands of samples, while certain minority classes have only dozens. To prevent prompt overfitting, a pool of various templates was used to construct the training samples. The fine-tuned model was then evaluated on the same test set.

\section{Results}
Our experiment results are as follows.

\subsection{Embedding Performance}
The performance of various multimodal embedding models is summarized in Table~\ref{tbl:embedding_results}. The \texttt{QQMM-embed} model~\citep{QQMM-embed} demonstrates the best clustering quality across the board. With enhanced class-definition prompting, it achieved a macro F1-score of 0.64 and an accuracy of 0.63. Without this prompting, its F1-score dropped to 0.55 and accuracy to 0.58. Generally, larger embedding models outperformed smaller ones, though they incur higher computational costs.

\begin{table*}[htbp]
  \centering
  \caption{Performance of five multimodal embedding models on the benchmark dataset. Arrows ($\downarrow, \uparrow$) indicate whether lower or higher values are preferable. QQMM performs better than GME~\citep{zhang2025gmeimprovinguniversalmultimodal}, mmE5~\citep{chen2025mme5improvingmultimodalmultilingual}, vdr~\citep{vdr}, and CLIP~\citep{DBLP:journals/corr/abs-2103-00020}.}
  \label{tbl:embedding_results}
  \footnotesize
  \begin{tabular}{l c c c c c c c c}
    \toprule
    \textbf{Model} & \textbf{Intra\,($\downarrow$)} & \textbf{Inter\,($\uparrow$)} & \textbf{Ratio\,($\uparrow$)} & \textbf{Silh.\,($\uparrow$)} & \textbf{DB\,($\downarrow$)} & \textbf{CH\,($\uparrow$)} & \textbf{F1\,($\uparrow$)} & \textbf{Acc.\,($\uparrow$)} \\
    \midrule
    \textbf{QQMM-embed} & 0.088 & 0.161 & 1.822 & 0.210 & 2.180 & 239.537 & 0.64 & 0.63 \\
    gme-Qwen2-VL-7b     & 0.128 & 0.098 & 0.761 & 0.074 & 3.361 &  95.294 & 0.59 & 0.62 \\
    mmE5-mllama-11b      & 0.143 & 0.112 & 0.785 & 0.089 & 3.286 &  95.304 & 0.51 & 0.53 \\
    vdr-2b-multi-v1      & 0.208 & 0.167 & 0.804 & 0.068 & 4.107 &  89.497 & 0.37 & 0.38 \\
    clip-ViT-L-14        & 0.205 & 0.110 & 0.536 & 0.002 & 5.231 &  65.732 & 0.18 & 0.22 \\
    \bottomrule
  \end{tabular}
\end{table*}

\subsection{VLM Performance}
VLMs demonstrated higher zero-shot classification accuracy. As shown in Table~\ref{tbl:vlm_results}, Qwen2.5-VL-72B achieved the best performance with a macro F1-score of 0.82 and an accuracy of 0.82. The advanced (``plus'') prompt provided a notable performance uplift over a simple (``base'') prompt for both 7B (10\% F1 lift) and 72B (5\% F1 lift) models. Qwen models outperformed other tested VLMs like Mistral Small 3.2~\citep{mistral} and Gemma 3~\citep{2025arXiv250319786G}.

\begin{table}[htbp]
  \centering
  \caption{Zero-shot classification performance of various VLMs. The ``plus'' prompt includes CoT and domain definitions while the ``base'' version mainly utilizes persona prompting.}
  \label{tbl:vlm_results}
  \begin{tabular}{l c c c c}
    \toprule
    \multirow{2}{*}{\textbf{Model}} & \multicolumn{2}{c}{\textbf{Base Prompt}} & \multicolumn{2}{c}{\textbf{Plus Prompt}} \\
    \cmidrule(lr){2-3} \cmidrule(lr){4-5}
    & Accuracy & F1 & Accuracy & F1 \\
    \midrule
    Qwen2.5-VL-72B  & 0.78 & 0.77 & 0.82 & 0.82 \\
    Qwen2.5-VL-7B   & 0.65 & 0.65 & 0.76 & 0.75 \\
    Gemma-3-27B      & 0.64 & 0.65 & 0.70 & 0.69 \\
    Mistral-3.2-24B  & 0.55 & 0.55 & 0.58 & 0.55 \\
    \bottomrule
  \end{tabular}
\end{table}

\subsection{SFT Performance}
Fine-tuning the Qwen2.5-VL-7B model yielded mixed results, i.e., the performance was dependent on the class distribution in the training data. Classes with thousands of training samples saw significant F1-score improvements (over 20\% uplift). Conversely, performance dropped for under-represented ones (which had dozens of training samples), highlighting the model's sensitivity to data imbalance. The model achieved 0.93 for both macro F1 and accuracy scores on a held-out test set for those classes over 150 training samples.

\section{Discussion and Conclusion}
\begin{figure*}[htbp]
  \centering
  \includegraphics[width=0.8\textwidth]{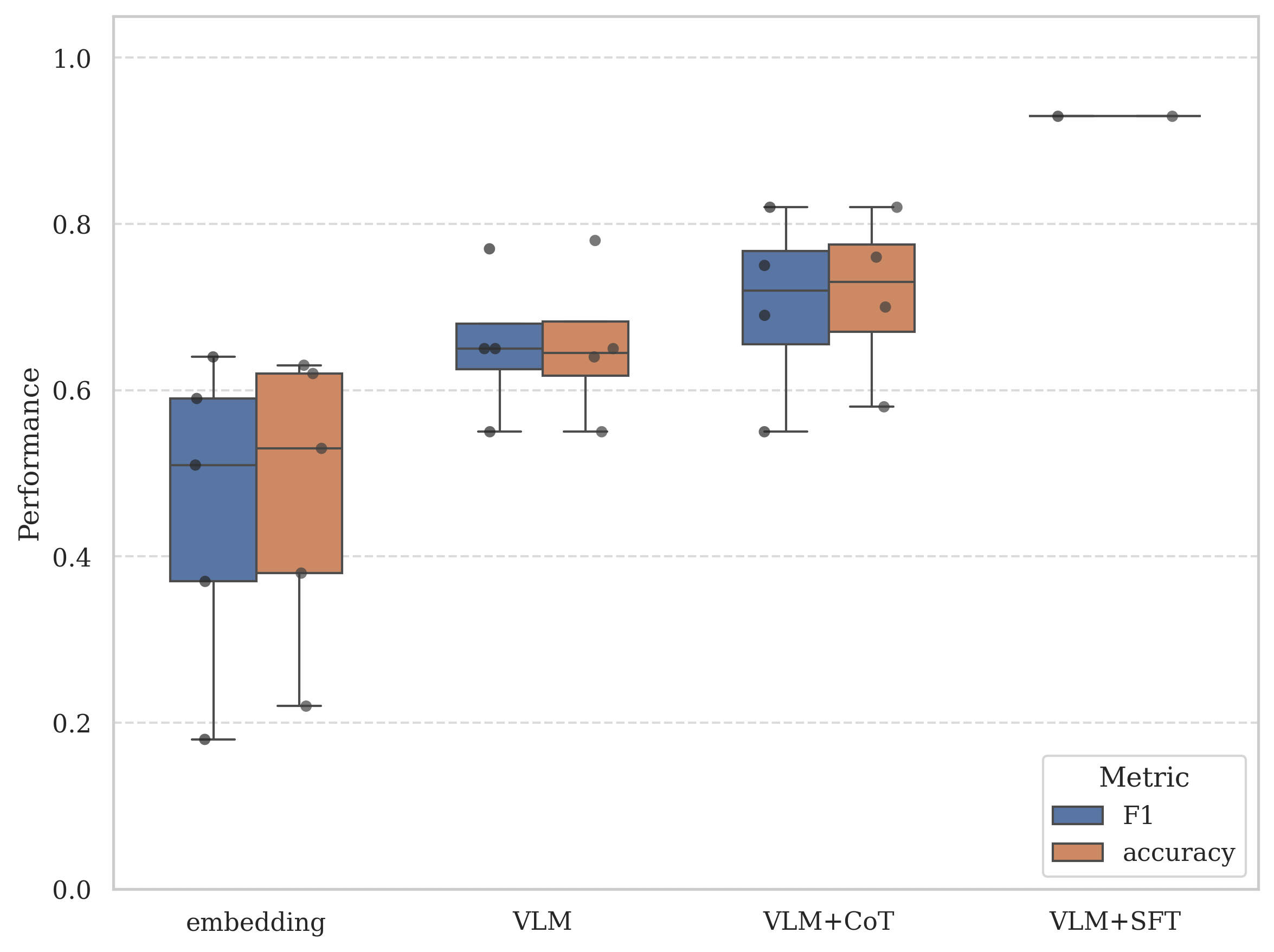}
  \caption{Classification performance for both embedding and VLMs under various configurations. VLMs generally outperform embedding models, especially after fine-tuned with sufficient training samples. Dots are individual models' results.}
  \label{fig:compare_all_clf}
\end{figure*}

A classification-focused comparison across all the choices is illustrated in Figure~\ref{fig:compare_all_clf}. VLMs' higher accuracy likely stems from their ability to perform deeper, end-to-end reasoning over the entire document image, capturing nuanced relationships between text and layout that are abstracted away into a single vector by embedding models. However, this comes at a cost. VLM inference is computationally expensive and slow, often requiring high-end GPUs to process large document images at scale. Furthermore, their generative nature can lead to nondeterministic outputs, a concern for production systems requiring reproducibility. In contrast, embedding models are lightweight, faster, and produce deterministic outputs, making them more suitable for large-scale batch processing on less powerful hardware.

For both model types, prompt engineering is a highly effective, low-cost method for injecting domain knowledge. For embedding models, providing detailed class definitions boosted F1-score from 0.55 to 0.64. For VLMs, the CoT-style ``plus'' prompt, which guides the model's reasoning process, lifted the 7B model's F1-score by 10 points. This underscores that off-the-shelf models are insufficient; performance is maximized when the models are guided by domain expertise. Successful domain adaptation through fine-tuning is achievable but requires a meticulous focus on creating well-balanced training datasets.

\bibliographystyle{plainnat}
\bibliography{example}

@inproceedings{10.5555/3600270.3602070,
author = {Wei, Jason and Wang, Xuezhi and Schuurmans, Dale and Bosma, Maarten and Ichter, Brian and Xia, Fei and Chi, Ed H. and Le, Quoc V. and Zhou, Denny},
title = {Chain-of-thought prompting elicits reasoning in large language models},
year = {2022},
isbn = {9781713871088},
publisher = {Curran Associates Inc.},
address = {Red Hook, NY, USA},
booktitle = {Proceedings of the 36th International Conference on Neural Information Processing Systems},
articleno = {1800},
numpages = {14},
location = {New Orleans, LA, USA},
series = {NIPS '22}
}

@article{DeBrabandere2017,
  author       = {Bert De Brabandere and
                  Davy Neven and
                  Luc Van Gool},
  title        = {Semantic Instance Segmentation with a Discriminative Loss Function},
  journal      = {CoRR},
  volume       = {abs/1708.02551},
  year         = {2017},
  url          = {http://arxiv.org/abs/1708.02551},
  eprinttype    = {arXiv},
  eprint       = {1708.02551},
  bibsource    = {dblp computer science bibliography, https://dblp.org}
}

@ARTICLE{QQMM-embed,
       author = {{Xue}, Youze and {Li}, Dian and {Liu}, Gang},
        title = "{Improve Multi-Modal Embedding Learning via Explicit Hard Negative Gradient Amplifying}",
      journal = {arXiv e-prints},
         year = 2025,
        month = may,
          eid = {arXiv:2506.02020},
        pages = {arXiv:2506.02020},
          doi = {10.48550/arXiv.2506.02020},
archivePrefix = {arXiv},
       eprint = {2506.02020},
 primaryClass = {cs.CV},
       adsurl = {https://ui.adsabs.harvard.edu/abs/2025arXiv250602020X}
}

@ARTICLE{zhang2025gmeimprovinguniversalmultimodal,
       author = {{Zhang}, Xin and {Zhang}, Yanzhao and {Xie}, Wen and {Li}, Mingxin and {Dai}, Ziqi and {Long}, Dingkun and {Xie}, Pengjun and {Zhang}, Meishan and {Li}, Wenjie and {Zhang}, Min},
        title = "{GME: Improving Universal Multimodal Retrieval by Multimodal LLMs}",
      journal = {arXiv e-prints},
         year = 2024,
        month = dec,
          eid = {arXiv:2412.16855},
        pages = {arXiv:2412.16855},
          doi = {10.48550/arXiv.2412.16855},
archivePrefix = {arXiv},
       eprint = {2412.16855},
 primaryClass = {cs.CL},
       adsurl = {https://ui.adsabs.harvard.edu/abs/2024arXiv241216855Z}
}

@inproceedings{chen2025mme5improvingmultimodalmultilingual,
    title = "mm{E}5: Improving Multimodal Multilingual Embeddings via High-quality Synthetic Data",
    author = "Chen, Haonan  and
      Wang, Liang  and
      Yang, Nan  and
      Zhu, Yutao  and
      Zhao, Ziliang  and
      Wei, Furu  and
      Dou, Zhicheng",
    editor = "Che, Wanxiang  and
      Nabende, Joyce  and
      Shutova, Ekaterina  and
      Pilehvar, Mohammad Taher",
    booktitle = "Findings of the Association for Computational Linguistics: ACL 2025",
    month = jul,
    year = "2025",
    address = "Vienna, Austria",
    publisher = "Association for Computational Linguistics",
    url = "https://aclanthology.org/2025.findings-acl.433/",
    pages = "8254--8275",
    ISBN = "979-8-89176-256-5"
}

@misc{vdr,
  author = {{LlamaIndex}},
  year = {2025},
  title = {Model Card for vdr-2b-multi-v1},
  howpublished = {https://huggingface.co/llamaindex/vdr-2b-multi-v1},
  note = {Accessed: 2025-07-07}
}

@article{DBLP:journals/corr/abs-2103-00020,
  author       = {Alec Radford and
                  Jong Wook Kim and
                  Chris Hallacy and
                  Aditya Ramesh and
                  Gabriel Goh and
                  Sandhini Agarwal and
                  Girish Sastry and
                  Amanda Askell and
                  Pamela Mishkin and
                  Jack Clark and
                  Gretchen Krueger and
                  Ilya Sutskever},
  title        = {Learning Transferable Visual Models From Natural Language Supervision},
  journal      = {CoRR},
  volume       = {abs/2103.00020},
  year         = {2021},
  url          = {https://arxiv.org/abs/2103.00020},
  eprinttype    = {arXiv},
  eprint       = {2103.00020},
  bibsource    = {dblp computer science bibliography, https://dblp.org}
}

@ARTICLE{2025arXiv250319786G,
       author = {{Gemma Team} and {Kamath}, Aishwarya and {Ferret}, Johan and {Pathak}, Shreya and {Vieillard}, Nino and {Merhej}, Ramona and {Perrin}, Sarah and {Matejovicova}, Tatiana and {Ram{\'e}}, Alexandre and {Rivi{\`e}re}, Morgane and {Rouillard}, Louis and {Mesnard}, Thomas and {Cideron}, Geoffrey and {Grill}, Jean-bastien and {Ramos}, Sabela and {Yvinec}, Edouard and {Casbon}, Michelle and {Pot}, Etienne and {Penchev}, Ivo and {Liu}, Ga{\"e}l and {Visin}, Francesco and {Kenealy}, Kathleen and {Beyer}, Lucas and {Zhai}, Xiaohai and {Tsitsulin}, Anton and {Busa-Fekete}, Robert and {Feng}, Alex and {Sachdeva}, Noveen and {Coleman}, Benjamin and {Gao}, Yi and {Mustafa}, Basil and {Barr}, Iain and {Parisotto}, Emilio and {Tian}, David and {Eyal}, Matan and {Cherry}, Colin and {Peter}, Jan-Thorsten and {Sinopalnikov}, Danila and {Bhupatiraju}, Surya and {Agarwal}, Rishabh and {Kazemi}, Mehran and {Malkin}, Dan and {Kumar}, Ravin and {Vilar}, David and {Brusilovsky}, Idan and {Luo}, Jiaming and {Steiner}, Andreas and {Friesen}, Abe and {Sharma}, Abhanshu and {Sharma}, Abheesht and {Mayrav Gilady}, Adi and {Goedeckemeyer}, Adrian and {Saade}, Alaa and {Feng}, Alex and {Kolesnikov}, Alexander and {Bendebury}, Alexei and {Abdagic}, Alvin and {Vadi}, Amit and {Gy{\"o}rgy}, Andr{\'a}s and {Susano Pinto}, Andr{\'e} and {Das}, Anil and {Bapna}, Ankur and {Miech}, Antoine and {Yang}, Antoine and {Paterson}, Antonia and {Shenoy}, Ashish and {Chakrabarti}, Ayan and {Piot}, Bilal and {Wu}, Bo and {Shahriari}, Bobak and {Petrini}, Bryce and {Chen}, Charlie and {Le Lan}, Charline and {Choquette-Choo}, Christopher A. and {Carey}, CJ and {Brick}, Cormac and {Deutsch}, Daniel and {Eisenbud}, Danielle and {Cattle}, Dee and {Cheng}, Derek and {Paparas}, Dimitris and {Shivakumar Sreepathihalli}, Divyashree and {Reid}, Doug and {Tran}, Dustin and {Zelle}, Dustin and {Noland}, Eric and {Huizenga}, Erwin and {Kharitonov}, Eugene and {Liu}, Frederick and {Amirkhanyan}, Gagik and {Cameron}, Glenn and {Hashemi}, Hadi and {Klimczak-Pluci{\'n}ska}, Hanna and {Singh}, Harman and {Mehta}, Harsh and {Tushar Lehri}, Harshal and {Hazimeh}, Hussein and {Ballantyne}, Ian and {Szpektor}, Idan and {Nardini}, Ivan and {Pouget-Abadie}, Jean and {Chan}, Jetha and {Stanton}, Joe and {Wieting}, John and {Lai}, Jonathan and {Orbay}, Jordi and {Fernandez}, Joseph and {Newlan}, Josh and {Ji}, Ju-yeong and {Singh}, Jyotinder and {Black}, Kat and {Yu}, Kathy and {Hui}, Kevin and {Vodrahalli}, Kiran and {Greff}, Klaus and {Qiu}, Linhai and {Valentine}, Marcella and {Coelho}, Marina and {Ritter}, Marvin and {Hoffman}, Matt and {Watson}, Matthew and {Chaturvedi}, Mayank and {Moynihan}, Michael and {Ma}, Min and {Babar}, Nabila and {Noy}, Natasha and {Byrd}, Nathan and {Roy}, Nick and {Momchev}, Nikola and {Chauhan}, Nilay and {Sachdeva}, Noveen and {Bunyan}, Oskar and {Botarda}, Pankil and {Caron}, Paul and {Kishan Rubenstein}, Paul and {Culliton}, Phil and {Schmid}, Philipp and {Sessa}, Pier Giuseppe and {Xu}, Pingmei and {Stanczyk}, Piotr and {Tafti}, Pouya and {Shivanna}, Rakesh and {Wu}, Renjie and {Pan}, Renke and {Rokni}, Reza and {Willoughby}, Rob and {Vallu}, Rohith and {Mullins}, Ryan and {Jerome}, Sammy and {Smoot}, Sara and {Girgin}, Sertan and {Iqbal}, Shariq and {Reddy}, Shashir and {Sheth}, Shruti and {Ptildeoder}, Siim and {Bhatnagar}, Sijal and {Raghuram Panyam}, Sindhu and {Eiger}, Sivan and {Zhang}, Susan and {Liu}, Tianqi and {Yacovone}, Trevor and {Liechty}, Tyler and {Kalra}, Uday and {Evci}, Utku and {Misra}, Vedant and {Roseberry}, Vincent and {Feinberg}, Vlad and {Kolesnikov}, Vlad and {Han}, Woohyun and {Kwon}, Woosuk and {Chen}, Xi and {Chow}, Yinlam and {Zhu}, Yuvein and {Wei}, Zichuan and {Egyed}, Zoltan and {Cotruta}, Victor and {Giang}, Minh and {Kirk}, Phoebe and {Rao}, Anand and {Black}, Kat and {Babar}, Nabila and {Lo}, Jessica and {Moreira}, Erica and {Martins}, Luiz Gustavo and {Sanseviero}, Omar and {Gonzalez}, Lucas and {Gleicher}, Zach and {Warkentin}, Tris and {Mirrokni}, Vahab and {Senter}, Evan and {Collins}, Eli and {Barral}, Joelle and {Ghahramani}, Zoubin and {Hadsell}, Raia and {Matias}, Yossi and {Sculley}, D. and {Petrov}, Slav and {Fiedel}, Noah and {Shazeer}, Noam and {Vinyals}, Oriol},
        title = "{Gemma 3 Technical Report}",
      journal = {arXiv e-prints},
         year = 2025,
        month = mar,
          eid = {arXiv:2503.19786},
        pages = {arXiv:2503.19786},
          doi = {10.48550/arXiv.2503.19786},
archivePrefix = {arXiv},
       eprint = {2503.19786},
 primaryClass = {cs.CL},
       adsurl = {https://ui.adsabs.harvard.edu/abs/2025arXiv250319786G}
}

@misc{mistral,
  author = {{Mistral AI}},
  year = {2025},
  title = {Model Card for Mistral-Small-3.2-24B-Instruct-2506},
  howpublished = {https://huggingface.co/mistralai/Mistral-Small-3.2-24B-Instruct-2506},
  note = {Accessed: 2025-07-07}
}

@ARTICLE{2025arXiv250213923B,
       author = {{Bai}, Shuai and {Chen}, Keqin and {Liu}, Xuejing and {Wang}, Jialin and {Ge}, Wenbin and {Song}, Sibo and {Dang}, Kai and {Wang}, Peng and {Wang}, Shijie and {Tang}, Jun and {Zhong}, Humen and {Zhu}, Yuanzhi and {Yang}, Mingkun and {Li}, Zhaohai and {Wan}, Jianqiang and {Wang}, Pengfei and {Ding}, Wei and {Fu}, Zheren and {Xu}, Yiheng and {Ye}, Jiabo and {Zhang}, Xi and {Xie}, Tianbao and {Cheng}, Zesen and {Zhang}, Hang and {Yang}, Zhibo and {Xu}, Haiyang and {Lin}, Junyang},
        title = "{Qwen2.5-VL Technical Report}",
      journal = {arXiv e-prints},
         year = 2025,
        month = feb,
          eid = {arXiv:2502.13923},
        pages = {arXiv:2502.13923},
          doi = {10.48550/arXiv.2502.13923},
archivePrefix = {arXiv},
       eprint = {2502.13923},
 primaryClass = {cs.CV},
       adsurl = {https://ui.adsabs.harvard.edu/abs/2025arXiv250213923B}
}

\end{document}